\documentclass[onecolumn,preprint]{aastex63}

\usepackage{amsmath}

\received{received date}
\revised{revised date}
\accepted{accepted date}
\submitjournal{ApJL}

\begin{document}
\title{Effect of Sea-ice Drift on the Onset of Snowball Climate on Rapidly Rotating Aqua-planets}
\author[0000-0002-4339-0489]{Wenshuo Yue,}
\affiliation{Department of Atmospheric and Oceanic Sciences, School of Physics, Peking University, Beijing 100871, China.}
\author[0000-0001-6031-2485]{Jun Yang}
\affiliation{Department of Atmospheric and Oceanic Sciences, School of Physics, Peking University, Beijing 100871, China.}
\correspondingauthor{Jun Yang}
\email{junyang@pku.edu.cn}

\begin{abstract}
Previous studies have shown that sea-ice drift effectively promote the onset of a globally ice-covered snowball climate for paleo Earth and for tidally locked planets around low-mass stars. Here, we investigate whether sea-ice drift can influence the stellar flux threshold for a snowball climate onset on rapidly rotating aqua-planets around a Sun-like star. Using a fully coupled atmosphere--land--ocean--sea-ice model with turning on or off sea-ice drift, a circular orbit with no eccentricity ($e$\,=\,0) and an eccentric orbit ($e$\,=\,0.2) are examined. When sea-ice drift is turned off, the stellar flux threshold for the snowball onset is 1250--1275 and 1173--1199~W\,m$^{-2}$ for $e$\,=\,0 and 0.2, respectively. The difference is mainly due to the poleward retreat of sea ice and snow edges when the planet is close to the perihelion in the eccentric orbit. When sea-ice drift is turned on, the respective stellar flux threshold is 1335--1350 and 1250--1276~W\,m$^{-2}$. These mean that sea-ice drift increases the snowball onset threshold by $\approx$80~W\,m$^{-2}$ for both $e$\,=\,0 and 0.2, promoting the formation of a snowball climate state. We further show that oceanic dynamics have a small effect, $\le$\,26~W\,m$^{-2}$, on the snowball onset threshold. This is because oceanic heat transport becomes weaker and weaker as the sea ice edge is approaching the equator. These results imply that sea-ice dynamics are important for the climate of planets close to the outer edge of the habitable zone, but \textcolor{
black}{oceanic heat transport is}\,\,less important.

\end{abstract}




\section{Introduction}\label{introduction}


Sea ice is a fundamental component of Earth climate system because it strongly influences the surface albedo and reduces the direct interactions between atmosphere and ocean \citep{curry1995sea,gardner2010a}. Sea ice also affects ocean stratification and thermohaline circulation because sea ice formation process causes expulsion of salt and heat into the ocean and sea ice melting process releases freshwater into the ocean and absorbs heat from the ocean and the overlying atmosphere \citep{warren1983why,holland2001the}. Driven by wind and ocean stresses, sea ice flows from the formation region to another region, which influences ice coverage, surface heat flux, and surface temperature, enhancing the role of sea ice in climate \citep{hibler1979a,thorndike,kimura2004sea,driftbook,Park&Stewart}.

Because of the important role of sea ice, how it influences the climate and habitability of planets beyond the solar system is a hot topic. Previous simulations of terrestrial exoplanets focus on the albedo effect of sea ice and snow \cite[e.g.,][]{shields2014spectrumdriven, Linsenmeier,Salamehetal2018}. These studies suggest that ice-albedo feedback is effective in promoting the onset of a snowball state when stellar flux or greenhouse gas concentration is low. And, the strength of the ice-albedo feedback is smaller on planets orbiting around low-mass stars due to the wavelength-dependent albedos of ice and snow \citep{warren1980b, wiscombe1980a} and to the \textcolor{black}{redder spectra of the host stars} \citep{joshi2012suppression, Shields2013}. In these studies, the sea ice is immobile because sea-ice dynamics are not included in their models. However, studies of paleo-Earth in 750--630 million years ago \citep{lewis2007snowball, yang2012first, voigt2012seaice, liu2018influence} and tidally locked planets around low-mass stars \citep{yang2019transition} have shown that sea-ice drift is also effective in promoting the onset of a snowball climate.

\cite{voigt2012seaice} showed that when sea-ice drift is considered, the CO$_2$ concentration threshold for a snowball Earth formation is $\approx$50--100 times higher (204--209 verus 2--4 ppmv), compared with the condition without sea-ice drift. The radiation forcing between these two levels of CO$_2$ is $\approx$20~W\,m$^{-2}$ (ref.~Figure~3 in \cite{ByrneandGoldblatt2014}), which is equivalent to a stellar flux of $\approx$114~W\,m$^{-2}$, i.e., 4$\times$25/(1-$\alpha_p$), where the factor of 4 is the ratio of planet's surface area to its cross-sectional area and $\alpha_p$ is the planetary albedo (0.3 is used in this simple estimate). \cite{yang2019transition} showed that \textcolor{black}{for an aqua-planet with no continent} sea-ice drift increases the stellar flux threshold for the onset of a tidally locked snowball state from 500--550 to 800--850~W\,m$^{-2}$, i.e., the effect of sea-ice drift is $\approx$300~W\,m$^{-2}$ (ref.~Supplementary Figure 3 in their paper). The different magnitudes for the effect of sea-ice drift in these two studies are mainly due to different continents (real continents versus aqua-planet) and different orbital configurations (rapidly rotating versus 1:1 tidally locked), as a result the ice drift speed in the latter is much larger than that in the former.

In this study, we explore the effect of sea-ice drift on the stellar flux threshold for the onset of a snowball climate on rapidly rotating aqua-planets orbiting around a Sun-like star. Two orbits are examined, one with a zero eccentricity ($e$) and one with $e$\,$=$\,$0.2$. The value of $e$ influences the ratio of stellar flux at periastron to that at apoastron ($(1+e)^2/(1-e)^2$) and the annual-mean stellar flux by $1/\sqrt{1-e^2}$, which are respectively 2.25 and \textcolor{black}{1.0206} under $e$\,$=$\,$0.2$. Exoplanet measurements showed a wide range of eccentricities from 0 to 0.97 for gas giant planets, but for lower-mass planets in the habitable zone have smaller eccentricities especially when the system has multiple planets \citep{limbach2015exoplanet, xie2016exoplanet, bolmont2016habitability}, so that we choose a moderate $e$ of 0.2.


Both energy balance models and general circulation models have been employed to examine the role of eccentricity in climate \citep{williamspollard2002, dressing2010habitable, spiegel2010generalized, armstrong2014effects, wang2014climate, Wangetal2017, Linsenmeier, Shieldsetal2016, bolmont2016habitability, way2017effects}. These studies found that a high eccentricity can cause large seasonal variability when the surface is covered by ice or land, but the high thermal inertia of ocean is effective in reducing the magnitude of the seasonal cycle  \citep[e.g.,][]{Linsenmeier}. Increasing the eccentricity widens the parameter space in which the planetary surface is able to maintain liquid water, at least during a part of one year for planets orbiting around the outer edge of the habitable zone \cite[e.g.,][]{dressing2010habitable}. However, none of their employed models had included the effect of oceanic or sea-ice dynamics except \cite{way2017effects}, in which ocean flows and sea-ice drifts are included but they have not analyzed their effects.


The structure of this paper is as follows. In Section 2 we introduce the model CCSM3 and explain the experimental design. In Section 3 we show the results: the effects of sea-ice drift, varying the eccentricity, and oceanic dynamics are addressed, respectively. Finally, we draw the summary in Section 4.



\section{Methods} \label{methods}
We use a fully coupled atmosphere--ocean model, the Community Climate System Model version 3 (CCSM3). The model includes four components (atmosphere, ocean, sea ice, and land) and a coupler. The atmosphere component, the Community Atmosphere Model version 3 (CAM3), is a global atmospheric general circulation model \citep{collins2004CAM3}. The ocean component uses the Parallel Ocean Program (POP) version 1.4.3, which is an oceanic general circulation model using the primitive equations and Boussineq approximation \citep{Smithetal2004}. The sea ice component employs the Community Sea Ice Model version 5 (CSIM5, \cite{briegleb2004scientific}). It includes a thermodynamics module to describe thermodynamic processes and an elastic–viscous–plastic representation for sea ice dynamics \citep{bitz1999an,hunke1997model,hunke2002model}. \textcolor{black}{The sea ice dynamics predict ice flows based on winds, ocean currents, and a model of the material strength of the ice within which an  elastic--viscous--plastic method is used to solve the nonlinear viscous--plastic ice rheology \citep[for details, please see][]{hunke1997model}}. The resolution of atmosphere and land components is T31 that has 96 grids in longitude and 48 grids in latitude with 26 vertical levels from the surface to about 3~hPa, and the resolution of ocean and sea ice components is gx3v5 that has 116 grids in longitude and 100 girds in latitude. The sea ice has five levels and the ocean has 25 levels from surface to $\approx$5000 m. Only the top 16 levels were used here because the surface is set to be an aqua-ocean with an uniform depth of 1~km and with no continent. Sea-ice drift equation is:

\begin{equation}
    \rho \frac{\partial \textbf{u}}{\partial t}=\boldsymbol{\tau}_a+\boldsymbol{\tau}_o-\rho f\textbf{k}\times \textbf{u}+\rho g \nabla H_o +\nabla \cdot \sigma_{ij},
\end{equation}

where $\rho$ is the mean density of sea ice and snow over ice, $\textbf{u}$ represents sea ice velocity, and $f$ is the Coriolis parameter, $g$ is gravity, $H_o$ is sea surface height, and  $\sigma_{ij}$ is internal stress tensor. The terms on right side are the air/ice stress, ocean/ice stress, Coriolis force, surface pressure gradient force, and internal ice stress, respectively.

In order to explore the effect of sea-ice drift, two types of experiment were run, turning on and turning off sea-ice drift, labelled as fully coupled  atmosphere--ocean--sea-ice and coupled atmosphere--ocean, respectively. To isolate the role of oceanic circulation, we further preformed corresponding experiments using the atmosphere component only (CAM3). CAM3 is coupled with a 50-m immobile ocean (with no oceanic heat transport) and a thermodynamic sea ice module, and sea-ice drift is not considered in the model. Comparisons between the atmosphere-only experiments and the coupled atmosphere--ocean experiments can exhibit the effect of oceanic dynamics. Likewise, comparisons between the coupled atmosphere--ocean experiments and the fully coupled atmosphere--ocean--sea-ice experiments can address the effect of sea-ice dynamics.

\begin{table*}\label{table111}
\caption{The stellar flux threshold for the onset of a snowball climate on Earth or on a rapidly rotating aqua-planet. $e$ is the orbital eccentricity, and $\beta$ is the planetary obliquity. All these studies employed the solar spectrum and concentrations of greenhouse gases similar to or somewhat higher than the pre-industrial level; the orbital period is 365 earth days; and the rotation period is one earth day. The snowball onset threshold is calculated based on annual-mean stellar flux for both circular and eccentric orbits.}

\centering
\begin{tabular}{l l c c c c l r}     
\hline\hline
Model & Continent & $e$ & $\beta$& Ocean dyn. & Ice drift & Threshold & Investigator\\
      &         &    &     &          &          & [W\,m$^{-2}$]         &   \\
\hline
 CCSM3 & modern & 0.0167 & 23.44$^{\circ}$ & yes & yes &  1224--1230  & \cite{yang2012first} \\
\hline
 CAM4 & Aqua & 0 & 23$^{\circ}$ & no & no & 1251--1265 & \cite{shields2014spectrumdriven} \\
 \hline
PlaSim & Aqua &0 & 0$^{\circ}$ & no & no & 1229--1297  & \cite{Linsenmeier} \\
 &  & 0.5 & 0$^{\circ}$  & no   & no  & 1103--1182 \\

\hline
 ECHAM6 & Aqua & 0 & 0$^{\circ}$ & no & no & 1225--1266 &  \cite{Salamehetal2018} \\
\hline
 CCSM3 & Aqua & 0 & 0$^{\circ}$  & yes  & yes & 1335--1350  & This study\\
            &  & 0 & 0$^{\circ}$ & yes   & no  & 1250--1275  &  \\
            &  & 0 & 0$^{\circ}$ & no    & no  & 1250--1275  &  \\
            &  & 0.2 & 0$^{\circ}$ & yes   & yes & 1250--1276  &  \\
            & & 0.2 & 0$^{\circ}$ & yes   & no  & 1173--1199  &  \\
            & & 0.2 & 0$^{\circ}$ & no    & no  & 1199--1225 &  \\
\hline
\end{tabular}
\end{table*}

Experimental designs are similar to those in \cite{yang2019ocean,yang2019transition}, but for a rapid rotation orbit rather than a tidally locked orbit. Planetary radius and gravity are the same as Earth; rotation period is one earth day and orbital period is 365 earth days; planetary obliquity ($\beta$) is set to be zero; and the solar spectrum is used. Two eccentricities are examined, 0.0 and 0.2. A larger eccentricity of 0.4 was also tested, but the model blew up due to too strong seasonal cycle. The surface pressure is 1.0~bar, dominated by N$_2$. Concentrations of CO$_2$, CH$_4$, and N$_2$O are set to be 300, 0.8, and 0.27 parts per million by volume (ppmv), respectively. For the visible band ($<$0.7\,$\mu$m), the snow albedo is 0.91 and the ice albedo is 0.68 if surface temperature is below $-$1\,$^\circ$C. For the near infrared band ($>$0.7\,$\mu$m), it is 0.63 for snow and 0.30 for sea ice \citep{collins2004CAM3}. Between $-$1\,$^{\circ}$C and 0\,$^{\circ}$C, the surface albedo decreases linearly with temperature \citep{yang2012first}. The albedo of open ocean is varied from 0.05 to 0.1 for different solar zenith angles and is uniform for all wavelengths. Note that the snowball onset threshold is sensitivity to the ice and snow albedos \citep{pierrehumbert2011climate,yang2012first,yang2012CCSM4,liu2017strong}, although this is beyond the scope of this study. Due to the long integration required for the equilibrium of the coupled ocean and to the limit of computation sources, we were not able to test a wide range of parameters such as planetary obliquity, continental configuration, or CO$_2$ concentration.


A series of stellar flux were run from $\approx$1100 to 1400~W\,m$^{-2}$ with a minimum interval of 15 or 25~W\,m$^{-2}$. In this paper, the stellar flux at the semi-major axis and the time-averaged stellar flux over one orbit are labelled as $S_0$ and $S_{a}$, respectively. Following \cite{voigt2012seaice} and \cite{yang2019transition}, the threshold for the onset of a snowball climate is defined as the maximum stellar flux at which the planet enters a snowball state when the simulation was started from an ice-free state or a partially ice-covered state under which the ice edge is away from the critical latitude ($\approx$20$^{\circ}$S(N)) for runaway glaciation. Each experiment was started from a rest state \textcolor{black}{(ocean velocity being zero everywhere, sea surface temperature of 280.15~K everywhere, and the ocean temperature decreasing with depth from 280.15~K at the sea surface to 275.15 K at the bottom)} with completely no sea ice or with a small area of high-latitude ice. In order to make sure the initial state will not influence the snowball onset threshold, we re-run some critical cases those close to runaway glaciation under different initial conditions (see Fig.~\ref{fig1_ice} below). Moreover, due to the strong cooling at the beginning of the simulations, the atmosphere component of the model becomes numerical instability frequently \textcolor{black}{in convection or precipitation module}, so that we have to attempt different initial states and different time steps (900, 600, or 300 s in the atmosphere component) in the simulations.


\section{results}\label{results}

\subsection{Effect of Sea-ice Drift}\label{sec-drift}
\begin{figure*}
    \centering
    \includegraphics[scale=0.9]{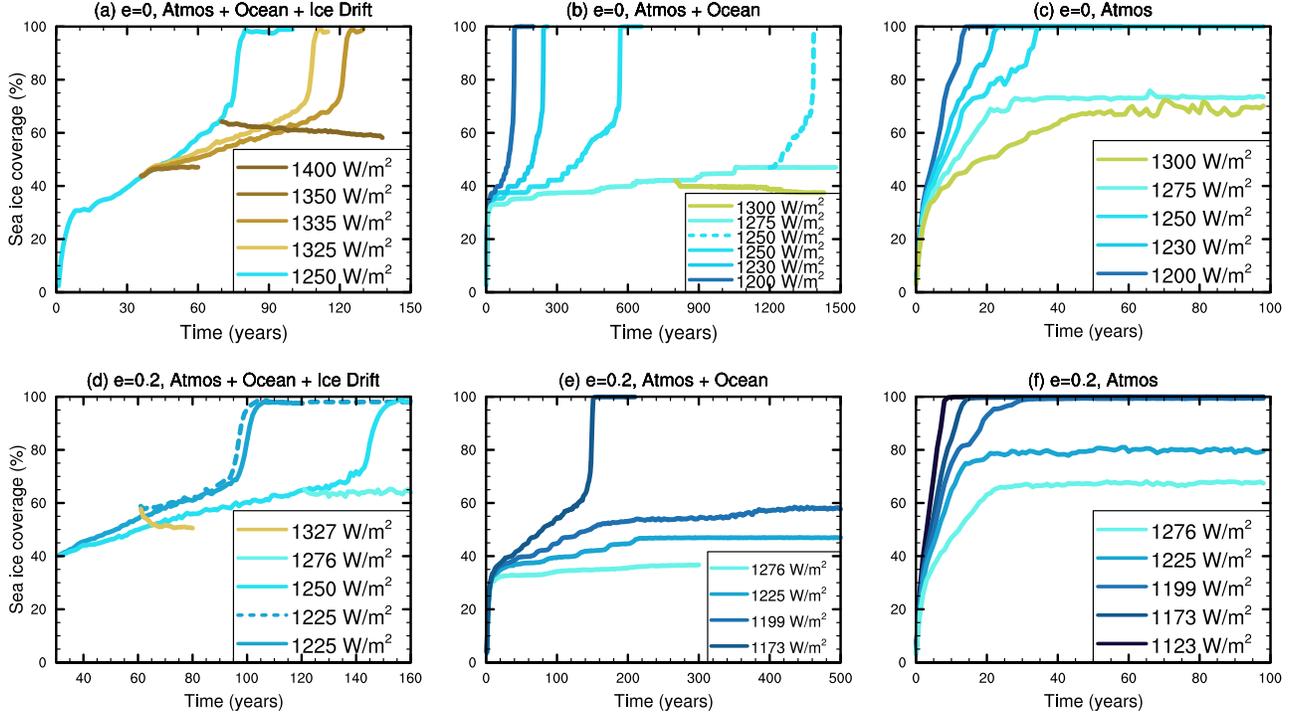}
    \caption{Evolution of annual- and global-mean sea ice coverage (\%) as varying the stellar radiation in the three groups of experiments. (a): $e$\,=\,0 in the fully coupled atmosphere--ocean--sea-ice experiments, (b): $e$\,=\,0 in the coupled atmosphere--ocean experiments without sea-ice drift, and (c): $e$\,=\,0 in the atmosphere-only experiments with neither oceanic dynamics nor sea-ice drift. (d--f): Same as (a--c) but for an eccentricity of 0.2. The figure legends are for annual-mean stellar fluxes for both $e$\,=\,0 and $e$\,=\,0.2. In the six panels, the stellar flux thresholds for the onset of a snowball climate are 1335--1350, 1250--1275, 1250--1275, 1250--1276, 1173--1199, and 1199--1225~W\,m$^{-2}$, respectively. The CO$_2$ concentration is 300~ppmv and the rotation period is one earth day. All the experiments were started from an ice-free state or a partial ice-covered state. Note that the cases of 1350 and 1400~W\,m$^{-2}$ in panel (a) and the case of 1327~W\,m$^{-2}$ in panel (d) had not reached exact equilibrium; unfortunately the model blew up due to numerical problem in convection. In (d), the cases of 1225 (solid line) and 1250~W\,m$^{-2}$ were restarted from the case of $e$\,=\,0 and 1250~W\,m$^{-2}$ in (a).}
    \label{fig1_ice}
\end{figure*}

Sea-ice drift acts to increase the stellar flux threshold for a snowball climate onset, promoting the snowball formation. As shown in Fig.~\ref{fig1_ice}(a-b) and summarized in Table~1, the threshold is 1335--1350 and 1250--1275 W\,m$^{-2}$ in the fully coupled atmosphere--ocean--sea-ice experiments (with sea-ice drift) and the coupled atmosphere--ocean experiments (without sea-ice drift), respectively. This means that sea-ice drift raises the stellar flux threshold by 75--85~W\,m$^{-2}$, i.e., 18.75--21.25~W\,m$^{-2}$ in global mean.

\begin{figure*}
    \centering
    \includegraphics[scale=1.0]{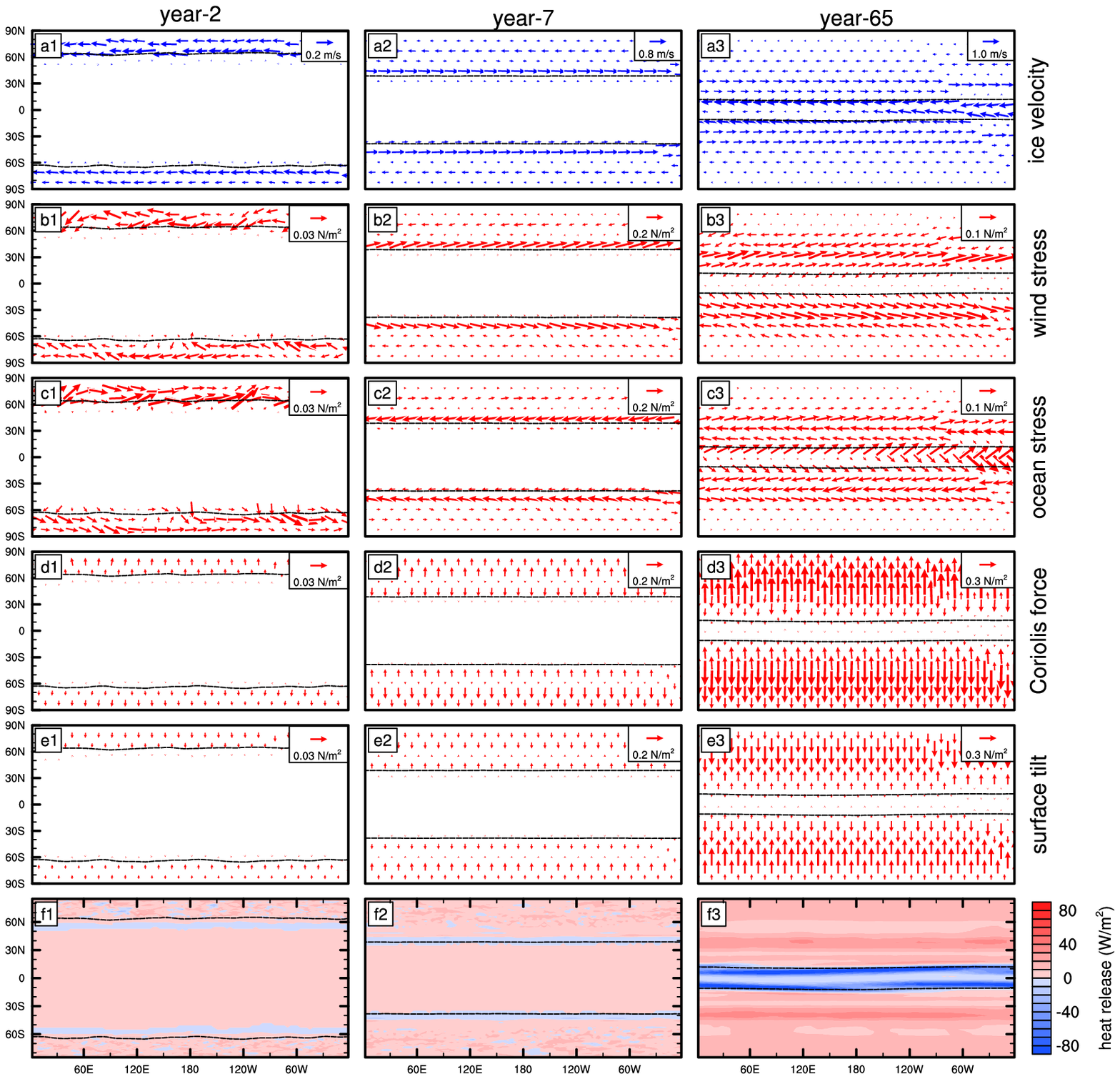}
    \caption{\textcolor{black}{Three different phases in the experiment of e=0 under a stellar flux of 1,250 W\,m$^{-2}$, started from an ice-free state. From left to right, the sea ice edges (marked using dashed lines) are at 60$^{\circ}$S(N), 40$^{\circ}$S(N), and 10$^{\circ}$S(N), corresponding to model years of 2, 7, and 65, respectively. (a1-a3): Sea ice velocity, (b1-b3): wind stress, (c1-c3): ocean stress, (d1-d3): Coriolis force, (e1-e3): sea surface tilt (i.e., the force due to the slope in sea surface elevation deviation away from the geoid), and (f1-f3): heat flux due to ice growth (positive) and ice melting (negative). The ice velocity and heat uptake become greater and greater as the ice edge is approaching the equator. Note the different reference vector lengths among panels.}}
    \label{fig2_drift}
\end{figure*}

The evolution of atmosphere, ocean, and sea ice in the atmosphere--ocean--sea-ice experiment of 1,250~W\,m$^{-2}$ in a circular orbit is shown in Fig.~\ref{fig2_drift}. \textcolor{black}{Due to the rapid rotation (same as Earth), the surface winds on each hemisphere are divided into three prevailing wind belts: polar easterlies from 60$^{\circ}$--90$^{\circ}$ latitude, prevailing westerlies between 30$^{\circ}$--60$^{\circ}$ latitude, and tropical easterlies within 30$^{\circ}$ latitude (i.e., trade winds). The Coriolis effect makes the wind curve to the west, to the east, and to the west, causing the winds blowing from the northeast toward the southwest, from southeast toward northeast, and from northeast toward the southwest, respectively \citep[see
chapter 1 in][]{WallaceandHobbs2006}. During the early phase for which the sea-ice edge is in the polar easterlies region (left panels), the equatorward movement of sea ice is mainly driven by surface winds while the ocean stresses and Coriolis forces have the opposite effect. Note that due to the absence of continents in the experiments, the surface winds are mainly in the zonal direction with a relatively smaller component in the meridional direction. The effect of sea surface tilts on sea ice (i.e., the force due to the slope in sea surface elevation deviation away from the geoid) is also equatorward, but its magnitude is much smaller than that of the surface wind stresses. When the sea ice edge has already entered the middle-latitude westerlies region between 30$^{\circ}$--60$^{\circ}$S(N) (middle panels), the equatorward expand of the sea ice near the ice edge is mainly driven by the combination of the wind stresses and Coriolis forces while the sea surface tilts exhibit the opposite effect. The ocean stresses also have an effect in expanding the ice pack to equatorward but with a smaller magnitude than that of the Coriolis forces. When the sea ice edge enters the tropical trade winds region (right panels), the equatorward drift of the ice mainly results from the surface winds. The ocean stresses act to resist the equatorward flow of sea ice, same as that shown in Fig.~13 of \cite{voigt2012seaice}. Around the ice edge, the effects of Coriolis forces and sea surface tilts are small, except in the center regions of the ice area their strengths are comparable to those of wind and ocean stresses.} During all the three phases, the magnitude of the internal ice stresses (not shown) is relatively small  \textcolor{black}{especially around the ice edge}.


Importantly, the speed of the sea-ice drift increases as the ice edge moves from the pole to the tropics, especially when the ice edge has already entered the region of tropical trade winds (black lines in Fig.~\ref{fig2_drift}(a1--a3)), promoting the onset of a snowball state. In the three phases, the meridional speed of the sea ice near the ice edges are $\approx$0.2, 0.5, and 1.0~m\,s$^{-1}$, respectively. Again, this means that the effect of sea-ice dynamics enhances when the ice edge is closer to the equator.

Another important process during the flow of sea ice is that a part of the ice melts when it meets with the relatively warm open-ocean water. This melting process absorbs energy from the ocean and the overlying atmosphere, cooling the surface and subsequently further promoting the equatorward expand of the ice. When the ice edge reaches the deep tropics, the heat uptake is as large as $\mathcal{O}$(80)~W\,m$^{-2}$ near the ice edge (Fig.~\ref{fig2_drift}(f3)). This mechanism is similar to that shown in \cite{yang2019transition} but for tidally locked planets.

Note that in the fully coupled atmosphere--ocean--sea-ice experiment of $e$\,=\,0, the snowball onset threshold is 1335--1350~W\,m$^{-2}$. It is about 100~W\,m$^{-2}$ higher than that in the study of \cite{yang2012first} that employed the same model. The difference is likely due to the fact that \cite{yang2012first} used a realistic continental configuration of modern Earth whereas an aqua-planet is used here. First, continents act to block or slow down the ice flows, reducing the effect of sea-ice dynamics \citep{yang2014water,yang2019transition}. Second, continents influence the hydrology cycle such that some central regions of large continents are snow-free even when the sea ice edge reaches the tropics, which constitutes a negative feedback due to the reduction of land surface albedo and thereby stabilizes the climate \citep{liu2018influence,Paradiseetal2019}. Moreover, the planetary obliquity is 23.44$^{\circ}$ and 0$^{\circ}$ in the previous study and this study, respectively; planets having greater obliquities can have ice-free areas at lower stellar fluxes \cite[e.g.,][]{Linsenmeier}. \textcolor{black}{The separate contributions of continent and obliquity are unknown yet.}


\subsection{Effect of Increasing the Eccentricity}\label{sec-eccentricity}

\begin{figure*}
    \centering
    \includegraphics[scale=1.0]{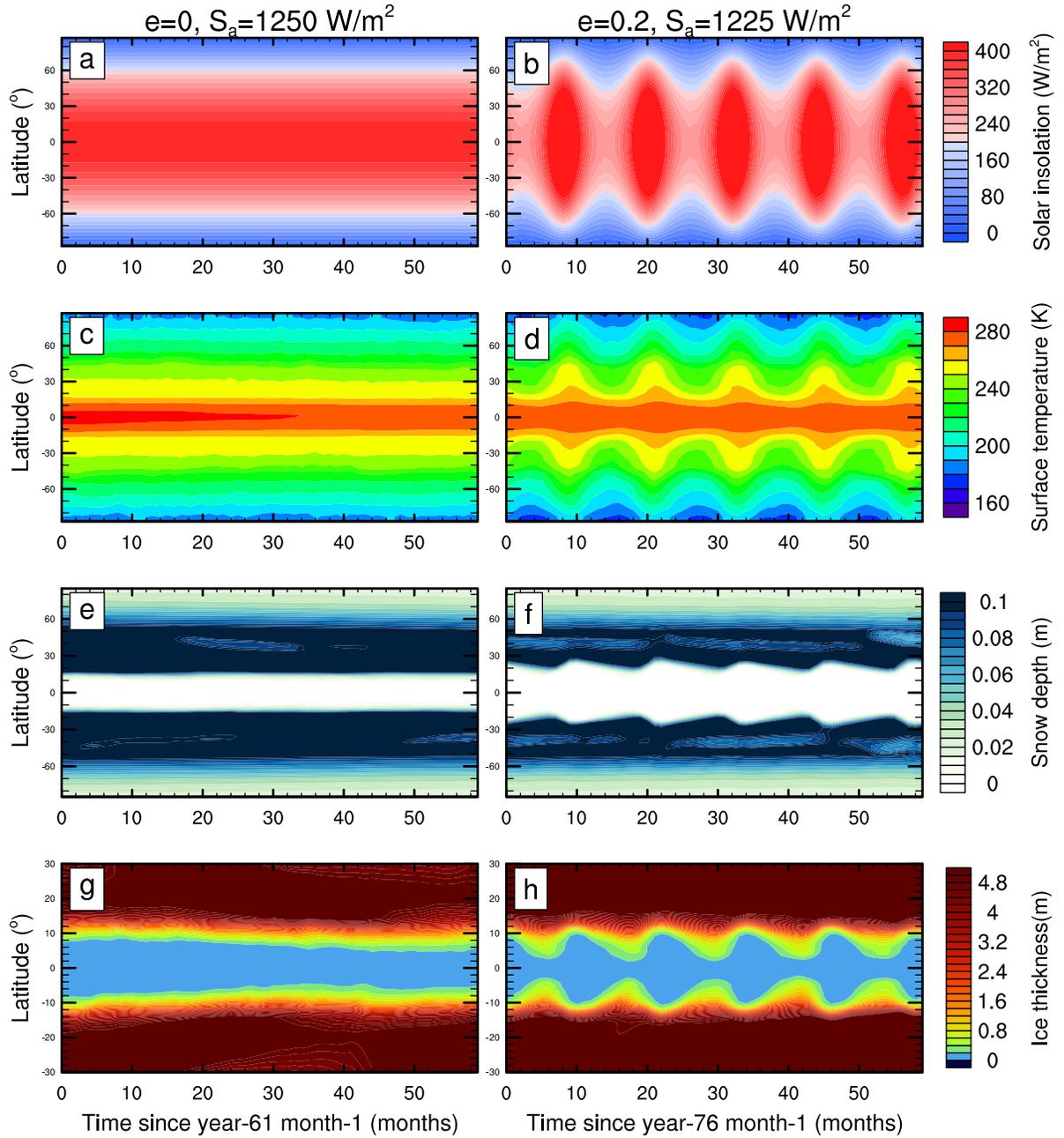}
    \caption{Comparisons in the evolution of stellar radiation (a--b), surface air temperature (c--d),  snow depth (e--f), and sea ice thickness (g--h) between $e$\,=\,0 and $e$\,=\,0.2. Left panels: for the fully coupled atmosphere--ocean--sea-ice experiment of $e$\,=\,0 under a  stellar flux of 1250 W\,m$^{-2}$, and right panels: for the fully coupled atmosphere--ocean--sea-ice experiment of $e$\,=\,0.2 under an annual-mean stellar flux of 1225 W\,m$^{-2}$.}
    \label{fig3_eccentricity}
\end{figure*}

The stellar flux threshold for the onset of a snowball climate decreases as increasing the eccentricity (Fig.~\ref{fig1_ice}). As listed in Table~1, when the eccentricity increases from 0.0 to 0.2, the annual-mean stellar flux thresholds for snowball onset decrease by 74--85, 76--77, and 50--51 W\,m$^{-2}$ in the fully coupled atmosphere--ocean--sea-ice, the coupled atmosphere--ocean, and the atmosphere-only experiments, respectively. The underlying mechanisms are shown in Fig.~\ref{fig3_eccentricity}. The higher stellar radiation around periastron on the orbit of $e$\,=\,0.2 trends to melt the surface ice and snow on ice, lowering the surface albedo and pushing the ice edge poleward. During the longer and colder winter near the orbit's apoastron, the surface temperatures decrease and the ice edge moves equatorward but the magnitudes are not large because the high thermal capacity of the ocean, similar to that addressed in \cite{dressing2010habitable}. Both ice thickness and snow depth near the ice edge in the case of $e$\,=\,0.2 are lower than those in the case of $e$\,=\,0 (Fig.~\ref{fig3_eccentricity}(e--h)), \textcolor{black}{despite the $e$\,=\,0.2 case has a relatively lower  annual-mean stellar flux, 1225 versus 1250~W\,m$^{-2}$}. As a result, the annual-mean surface albedo is lower and more stellar flux is absorbed at the surface in the eccentric orbit, delaying the onset of runaway glaciation and the formation of a snowball climate.



\subsection{Effect of Oceanic Dynamics}\label{sec_ocean}

Comparing the results of the coupled atmosphere--ocean experiments with those of the atmosphere-only experiments, Fig.~\ref{fig1_ice}(b) versus~\ref{fig1_ice}(c) and Fig.~\ref{fig1_ice}(e) versus~\ref{fig1_ice}(f), one can find the effect of oceanic dynamics. Under $e$\,=\,0, the stellar flux threshold is 1250--1275~W\,m$^{-2}$ for both types of experiments. Under $e$\,=\,0.2, the stellar flux threshold is 1173--1199~W\,m$^{-2}$ in the coupled atmosphere--ocean experiment and 1199--1225~W\,m$^{-2}$ in the atmosphere-only experiment. These results suggest that oceanic heat transport has a small (0--26~W\,m$^{-2}$) effect on the snowball onset threshold. In the previous studies of \cite{voigt2012seaice} and \cite{yang2019transition}, they also found a small effect of oceanic dynamics on the threshold. \cite{voigt2012seaice} showed that oceanic dynamics decrease the threshold from 4 to 2 ppmv in the concentration of CO$_2$ (ref. their Fig.~12). \cite{yang2019transition} showed that in both types of experiment the threshold is 500--550~W\,m$^{-2}$ for tidally locked aqua-planets (ref. their Fig.~S3). This is on the contrary to intuition as oceanic dynamics are able to transport heat from the tropics to the sea ice edge and melt the ice.

\begin{figure*}
    \centering
    \includegraphics[scale=0.8]{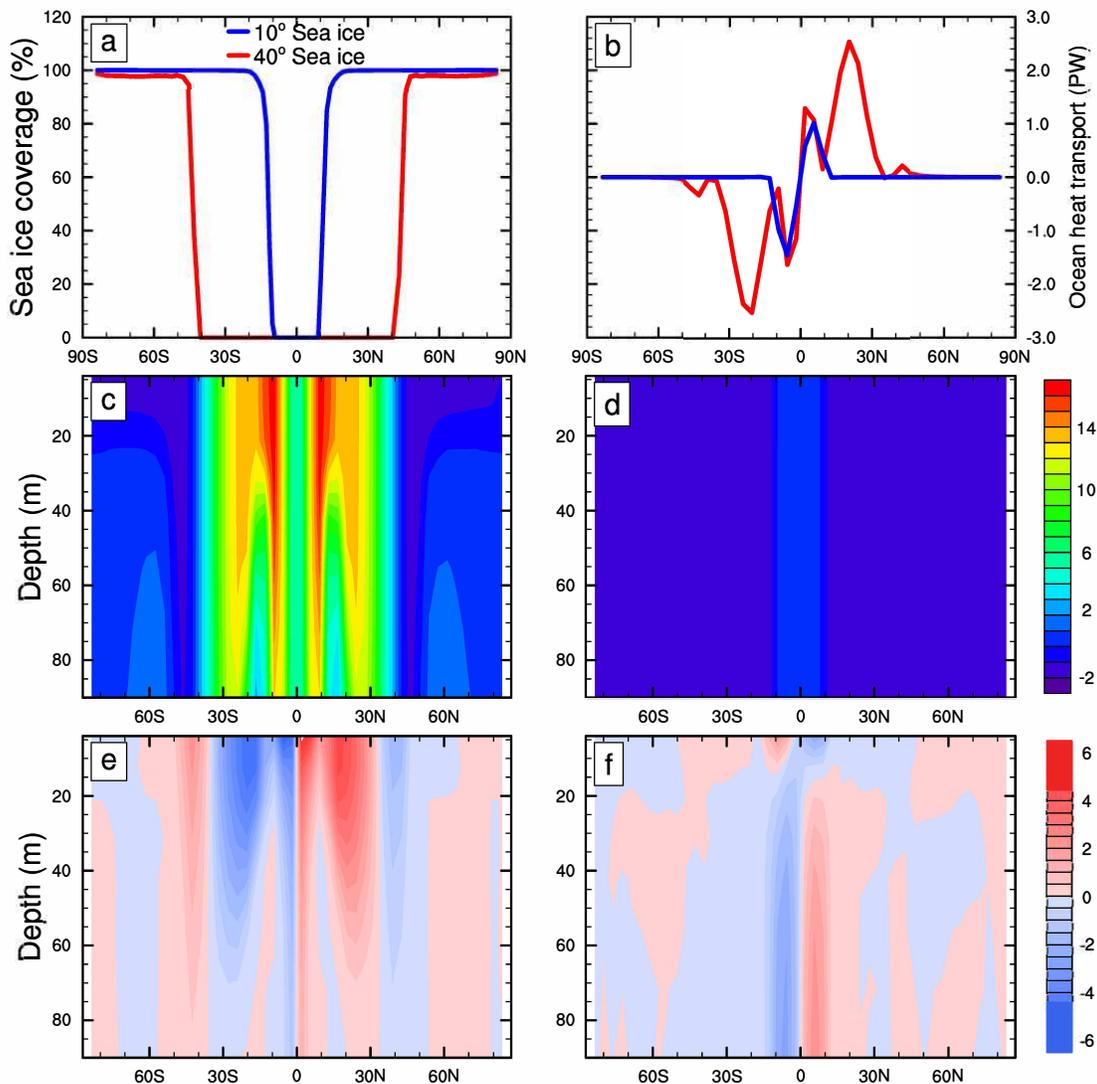}
    \caption{Effects of oceanic dynamics on the climate. (a) sea ice coverage in a climate when the ice edge is at around 40$^{\circ}$S(N) (red line) and in a relatively cool climate when the ice edge is at around 10$^{\circ}$S(N); (b) the corresponding meridional oceanic heat transports (PW, 1 PW\,=\,10$^{15}$~W); (c) ocean potential temperature in the relatively warm climate; (d) same as (c) but in the relatively cool climate; (e) meridional oceanic velocity in the relatively warm climate; and (f) same as (e) but in the relatively cool climate. These data are from the coupled atmosphere--ocean experiment of $e$\,=\,0 under a stellar flux of 1250 W\,m$^{-2}$. The ocean is 1000~m, but only the top 100~m is shown in (c-f); below the level of 100~m, ocean properties are much uniform and ocean currents are much weaker. In (c), equatorial temperatures are lower than those in higher latitudes; this is due to the effect of equatorial ocean upwelling driven by trade winds. The oceanic heat transport becomes weaker as the ice edge is closer to the equator because ocean currents weaken (especially near the sea surface) and oceanic temperature gradients become smaller.}
    \label{fig4_ocean}
\end{figure*}

The reason is that the magnitude of oceanic heat transport becomes smaller and smaller as the sea-ice edge is approaching the equator. Figure~\ref{fig4_ocean} shows the sea ice coverage, oceanic temperature, meridional ocean velocity, and meridional oceanic heat transport in two transient phases in the coupled atmosphere--ocean experiment of $e$\,=\,0 and S$_a$\,=\,1250~W\,m$^{-2}$; one phase is when the sea-ice edge is at around 40$^{\circ}$S(N) and the other one is when the sea-ice edge is at around 10$^{\circ}$S(N). It is clear that the ocean becomes cooler (Fig.~\ref{fig4_ocean}(c--d)) and the merdional ocean velocity becomes weaker (Fig.~\ref{fig4_ocean}(e--f)) when the sea ice is closer to the equator. Moreover, the meridional temperature gradients in the ocean decrease greatly when more surface is covered by ice and snow (Fig.~\ref{fig4_ocean}(d)). As a result, the oceanic heat transport decreases greatly (Fig.~\ref{fig4_ocean}(b)). In these two phases, the oceanic heat transport of the latter is only 15\% of the former. Therefore,  oceanic dynamics have a very small effect on the stellar flux threshold for the onset of a snowball state.

The relatively strong effect of oceanic dynamics when the ice edge is far away from the equator can been also found in the equilibrium states of experiments with relatively high stellar fluxes. As shown in Fig.~\ref{fig1_ice}(b--c), under the same eccentricity and the same stellar flux ($e$\,=\,0 and S$_a$\,=\,1275~W\,m$^{-2}$), the global-mean sea ice coverage is $\approx$47\% and $\approx$64\% in the coupled atmosphere--ocean and atmosphere-only experiments, respectively. Another example is that in the eccentric orbit of $e$\,=\,0.2 and S$_a$\,=\,1225~W\,m$^{-2}$, the respective global-mean sea ice coverage is $\approx$46\% and $\approx$80\% (Fig.~\ref{fig1_ice}(e--f)).

Note that in our atmosphere-only aqua-planet experiments with $e$\,=\,0 and $\beta$\,=\,0$^{\circ}$, the planet enters a snowball state when the stellar flux is $\approx$1250--1275~W\,m$^{-2}$. This value is similar to those in previous studies which also used atmosphere-only models and under the same eccentricity, such as 1251--1265~W\,m$^{-2}$ obtained in the model CAM4 but with a value of $\beta$\,=\,23$^{\circ}$ \citep{shields2014spectrumdriven}, 1229--1297~W\,m$^{-2}$ obtained in the model PlaSim with $\beta$\,=\,0$^{\circ}$ \citep{Linsenmeier}, and 1225--1266~W\,m$^{-2}$ in the model ECHAM6 with $\beta$\,=\,0$^{\circ}$ \citep{Salamehetal2018}. Differences in the threshold can be due to different parameters used in the models, such as ice and snow albedos and convection and cloud parameterizations.


\section{Summary}\label{summary}


The global climate model CCSM3 are employed to investigate the onset of a globally ice-covered snowball state on rapidly rotating planets around Sun-like stars. Three types of experiments were performed, atmosphere-only, coupled atmosphere--ocean, and fully coupled atmosphere--ocean--sea-ice. Three main conclusions are obtained as follows:

   \begin{enumerate}
    \renewcommand{\labelenumi}{(\theenumi)}
      \item Sea-ice drift promotes the onset of a snowball climate state. It increases the stellar flux threshold for a snowball onset by $\approx$80~W\,m$^{-2}$ for rapidly rotating aqua-planets under both $e$\,=\,0 and 0.2. The underlying mechanisms are (i) sea ice flows from growth region to open-ocean region, increasing ice coverage and surface albedo, and (ii) a part of the ice melts when it flows to the relatively warm tropical ocean, during this process a significant amount of heat is absorbed from the ocean and the overlying atmosphere and subsequently the surface temperature decreases, further promoting the expand of sea ice.

      \item The climate in an eccentric orbit is warmer than its circular-orbit equivalent, as a result, the snowball onset threshold under $e$\,=\,0.2 is $\approx$50--85~W\,m$^{-2}$ (in annual-mean stellar flux) lower than that under $e$\,=\,0. The key mechanism is that surface ice and snow melt when the planet orbits close to the orbit's periastron, reducing the surface albedo. This conclusion is consistent with the previous studies of  \textcolor{black}{\cite{dressing2010habitable} and \cite{Linsenmeier}}.

      \item In general,\,\, \textcolor{black}{oceanic heat transport has} a warming effect on the climate. But, \textcolor{black}{it has}\,\, a small effect ($\le$\,26~W\,m$^{-2}$) on the snowball onset threshold because ocean temperature gradients and oceanic currents become weaker and weaker when the sea-ice edge is approaching the equator. \textcolor{black}{However, when the ice edge has already entered the deep tropics, ocean stress on the sea ice is effective in slowing down the equatorward spread of the sea ice, resisting the accelerating effect of wind stress, as firstly pointed out in \cite{voigt2012seaice}.}

   \end{enumerate}

Our results imply that sea-ice drift can affect the location of the outer edge of the habitable zone. In order to quantify the degree of this influence, future simulations using climate models those include massive CO$_2$ and CO$_2$ condensation are required; the model we employed here has no this capability yet. The flows of ocean and sea ice are mainly driven by surface winds, the strength of which depends on surface temperature gradient as well as surface pressure. The surface pressure is 1.0~bar in this study and future works with different surface pressures are required. \textcolor{black}{The strength of surface wind stress depends on surface wind speed and air density. For a more-massive atmosphere, the surface wind speed decreases due to more effective heat advections and thereby the surface temperature gradients reduce, however, the surface air density increases. Recent coupled atmosphere-ocean simulations of \cite{Olsonetal2020} showed that the net result is that the wind stress increases with air mass (Fig. 6d in their paper) although the increasing rate is much smaller than that expected from the increasing of surface air density. For example, when the surface pressure is increased from 1 to 10 bar, the surface wind stress nearly doubles. However, their experiments are for varying N$_2$ partial pressure rather than CO$_2$ partial pressure; future work is required for varying CO$_2$.} No continent is involved in our experiments. If continents were included, the effects of both oceanic heat transport and sea-ice drift will likely be weaker due to their friction and barrier effects.


\acknowledgments

We are grateful to Yaoxuan Zeng and Yonggang Liu for technical help. J.Y. acknowledges support from the National Natural Science Foundation of China (NSFC) grants 41861124002 and 41675071.



\bibliographystyle{aasjournal}
\bibliography{ref}





\end{document}